\documentclass[aps,prl,twocolumn,superscriptaddress,nofootinbib]{revtex4-1}

\usepackage{hyperref}
\usepackage{color}
\usepackage{graphicx,psfrag, subfigure}
\usepackage{color,soul}
\usepackage{tensor}

\usepackage{mathtools}
\usepackage{braket}

\usepackage{array}

\usepackage{amssymb}
\usepackage{amsmath}
\DeclareSymbolFont{bbold}{U}{bbold}{m}{n}
\DeclareSymbolFontAlphabet{\mathbbold}{bbold}
\usepackage{color}
\usepackage{eufrak}
\usepackage{graphicx,psfrag, subfigure}
\usepackage{hyperref}
\DeclareMathAlphabet{\mathpzc}{OT1}{pzc}{m}{it}
\usepackage{blkarray}

\usepackage{yfonts}
\usepackage{mathrsfs}
\usepackage{relsize}
\usepackage{tabularx}

\newcommand{\diag}{\text{diag}}

\DeclareMathAlphabet\mathbfcal{OMS}{cmsy}{b}{n}

\def\Qp{{\fontfamily{cmbx10}\selectfont Q}}
\def\Q{\textnormal \Qp}

\def\M{M_{{\rm{pl}}}}

\def\be{\begin{equation}}
\def\ee{\end{equation}}
\def\bea{\begin{eqnarray}}
\def\eea{\end{eqnarray}}

\def\be{\begin{equation}}
\def\ee{\end{equation}}
\def\bea{\begin{eqnarray}}
\def\eea{\end{eqnarray}}
\usepackage{dsfont}
\usepackage[mathscr]{eucal}

\newcommand{\bs}{\boldsymbol}

\def\simleq{\; \raise0.3ex\hbox{$<$\kern-0.75em
      \raise-1.1ex\hbox{$\sim$}}\; }

\def\simgeq{\; \raise0.3ex\hbox{$>$\kern-0.75em
      \raise-1.1ex\hbox{$\sim$}}\; }

\begin{document}

\title{Axions of Evil}

\author{Thomas C. Bachlechner}
\author{Kate Eckerle}
\affiliation{Department of Physics, Columbia University, New York, USA}
\author{Oliver Janssen}
\author{Matthew Kleban}
\affiliation{Center for Cosmology and Particle Physics, New York University, New York, USA and New York University Abu Dhabi, Abu Dhabi, UAE}
\vskip 4pt

\begin{abstract}

{We provide a systematic framework for theories of multiple axions.  We discover a novel type of ``alignment'' that renders even very complex theories analytically tractable. Theories with $\sim 100$ axions and random parameters have an exponential number of meta-stable vacua and accommodate a diverse range of inflationary observables.  Very light fields can serve as dark matter with the correct abundance.
Tunneling from a minimum with large vacuum energy can occur via a thin-wall instanton and be followed by a sufficient period of slow-roll inflation that ends in a vacuum containing axion dark matter and  a cosmological constant with a value consistent with observation.  Hence, this model can reproduce many macroscopic features of our universe without   tuned  parameters.}

\end{abstract}

\maketitle

\section{Introduction}

Axionic fields arise in a number of contexts \cite{Peccei:1977hh, Svrcek:2006yi, Natural}.  Like quantized fluxes \cite{Bousso:2000xa}, theories of multiple axions \cite{Bachlechner:2015gwa} can naturally accommodate the observed cosmological constant.    Furthermore, the unbroken discrete shift symmetry of axions  provides  a natural  inflaton candidate \cite{Natural,Nflation,KNP,McAllister:2008hb,Kaloper:2008fb,Kaloper:2011jz}, and the absence of direct detection and the issues with  conventional dark matter models at sub-galaxy length scales have led to a recent surge of interest in ultralight axions as dark matter \cite{Hu:2000ke, Arvanitaki:2009fg, Hui:2016ltb,Marsh:2015xka,Diez-Tejedor:2017ivd}.  In this note we  demonstrate that generic theories of multiple axions with a single energy scale near the fundamental scale can simultaneously fulfill all three roles.  These theories also provide high energy meta-stable vacua that can serve as an initial or generic state for the universe.

Axions are protected by continuous shift symmetries $\theta^i\rightarrow \theta^i+c^i$, all of which are broken by nonperturbative effects at scales  $\Lambda_I^4\sim e^{-S_I}\Lambda_{\text{UV}}^4$, where $S_I$ denotes the action of the $I^{\text{th}}$ among $P$ leading instantons. The axions couple to each instanton through integer charges ${\mathbfcal Q}^I$,\footnote{Bold font denotes field space vectors and matrices, e.g. ${\mathbfcal Q}^I$ is a row vector, and we set $\hbar = 1$.} resulting in a nonperturbative potential of the form
\be\label{nppotentialtheta}
V_{\text{NP}}=\sum_{I=1}^P \Lambda_I^4 \left[1-\cos\left({\mathbfcal Q}^I\, \bs\theta+\delta^I\right)  \right]+~\dots\, 
\ee
where the phases of the instantons are denoted by $\delta^I$ and ellipses represent subleading terms. The simple form of the nonperturbative potential renders axion models amenable to concrete computations.  A main result of our work is the efficient identification of approximate symmetries of $V_{\text{NP}}$, which allows for a systematic description of the  physics. In this note we present some key results, while further details will be published elsewhere \cite{bk,bejk2}.
\begin{figure}[h]
\centering
\includegraphics[width=.48\textwidth]{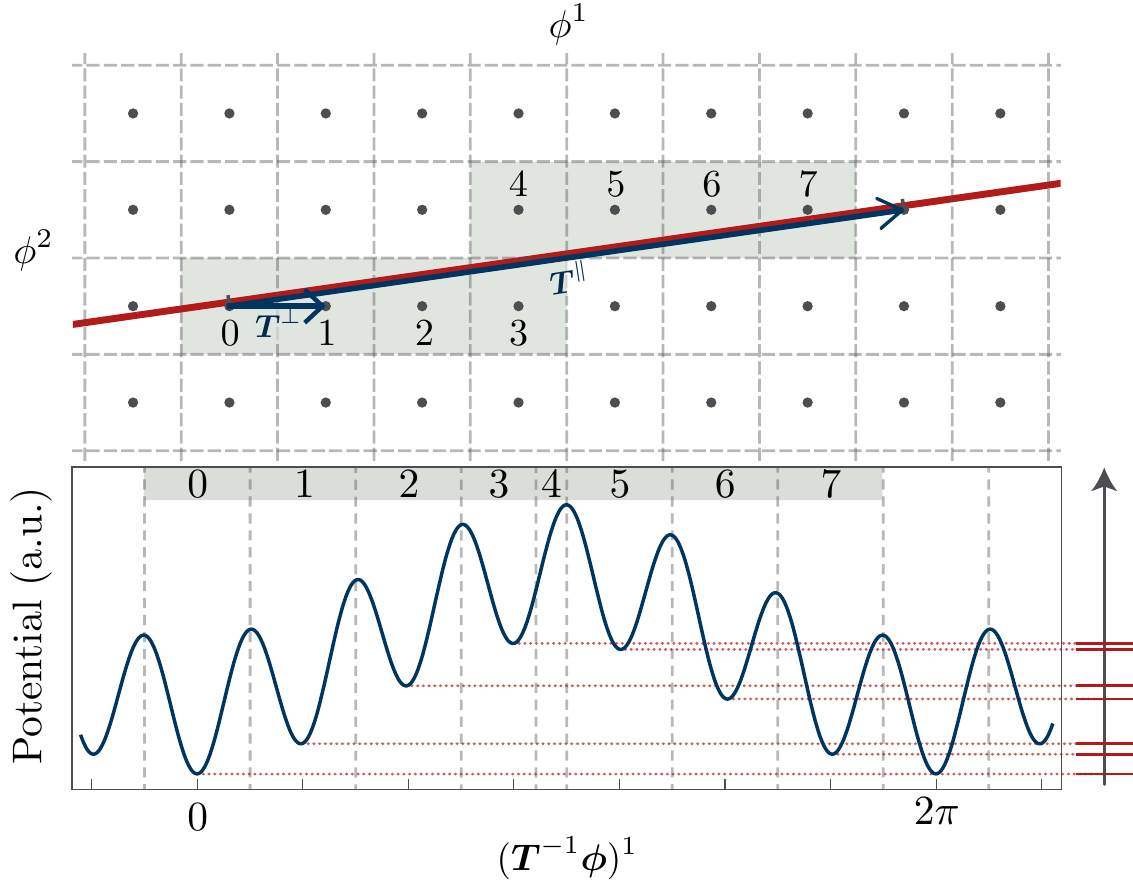}
\caption{\small Top: Constraint surface along with the lattice $2\pi \mathbb Z^P$ (gray dots). Arrows show the aligned basis vectors $\bs T^\parallel$ and $\bs T^\perp$. Bottom: Axion potential. Distinct fundamental domains are numbered and shaded.}\label{fig:potplot}
\end{figure}

\section{Alignment}

Consider a theory with $N$ axions $\theta^{i}$,
\be\label{lagtheta}
{\cal L}={1\over 2}\partial \bs\theta^\top\bs K \partial \bs\theta-V_{\text{NP}}(\bs \theta)-V_0\,,
\ee
where $\bs K$ is the field space metric and $V_0$ denotes a background vacuum energy density.

The leading axion potential in (\ref{nppotentialtheta}) is invariant under the $P$ identifications ${\mathbfcal Q}^I\, \bs\theta \cong {\mathbfcal Q}^I\, \bs\theta + 2\pi$. To take full advantage of this symmetry we define the {\it  lattice basis} by promoting all $P$ cosine arguments to $P$ fields $\phi^I$ that are constrained to reproduce (\ref{nppotentialtheta}),
\be \label{phibasis}
\bs\phi \simeq {\mathbfcal Q} \, \bs \theta \, + \, \bs \delta\,,~~~{\mathbfcal Q} = \left(\begin{matrix} \mathbf Q\\\mathbf Q_{\text{R}}\end{matrix}\right)\,,
\ee
where $\simeq$ denotes equality after imposing  equations of motion, and we decomposed the charges into an invertible $N \times N$ matrix $\mathbf Q$ and a rectangular matrix $\mathbf Q_\text{R}$ containing the $P-N$ remaining charge vectors. The potential becomes
\be \label{phipot}
V_{\text{NP}}=\begin{underbrace} {\sum_{I=1}^P \Lambda_I^4 \left[1-\cos(\phi^I ) \right]}\end{underbrace}_{\equiv V_{\text{aux}}}+\sum_{j=1}^{P-N} \nu_j \mathbfcal R^j \left( \bs \phi -\bs \delta \right)\,,
\ee
where 
$$\mathbfcal R=\left(\begin{matrix}\mathbf Q_\text{R} \mathbf Q^{-1} \, \,  \vert \, \,  -\mathbbold{1}_{P-N} \end{matrix}\right)$$ 
is a $P-N \times P$ matrix and $\nu_j$ are  Lagrange multipliers that fix $P-N$ of the fields so as to reproduce \eqref{nppotentialtheta}. 

The potential is now $V_{\text{aux}}$ evaluated on the $N$-dimensional surface  
$$\mathbfcal R( \bs \phi -\bs \delta ) = \bs 0.$$ 
The constraint surface, illustrated in Figure \ref{fig:potplot}, slices through multiple distinct fundamental domains of the simple integer lattice $2\pi \mathbb Z^P$, on which $V_{\text{aux}}$ is periodic.  Each domain is labeled by an integer vector $\bs n$ and defined by\footnote{The $L^\infty$-norm  is defined as $\lVert \bs v \rVert_\infty \equiv {\text{max}}_i \{ |v^i| \}$.}
\be\label{funddomain}
\lVert \bs \phi - 2\pi \bs n \rVert_\infty \le\pi\, .
\ee

The introduction of auxiliary fields now allows us to easily identify both exact and approximate periodicities of the on-shell potential and represents a key part of our work.
To exhibit these symmetries we define an {\it aligned basis}, $\bs T$. This basis is chosen so that it minimizes the projections of the basis vectors onto the orthogonal complement of the constraint surface. More precisely, $\mathbfcal P\bs T$ is a reduced basis  of the lattice generated by $\mathbfcal P$, containing the shortest vectors under the $L^\infty$-norm, where
\be \label{projectedbasis}
\mathbfcal P \equiv  \mathbfcal R^\top(\mathbfcal R \mathbfcal R^\top)^{-1}\mathbfcal R\,,
\ee
such that $\mathbbold{1}_{P} - \mathbfcal P$  is the orthogonal projector onto the constraint surface.\footnote{In practice the basis $\boldsymbol T$ may be hard to find, but an approximation is easily obtained via the extended LLL algorithm \cite{Lenstra1982}, which finds a basis reduced under the $L^2$-norm, see for example  \cite{lllpackage}.}

We can decompose the basis $\bs T$ into $N$ vectors parallel to the constraint surface, $\mathbfcal R\bs T^\parallel_{i}= \bs 0$, and $P-N$ vectors generating non-vanishing translations transverse to the constraint surface, $\mathbfcal R\bs T^\perp_{j }\ne \bs 0$, as shown in Figure \ref{fig:potplot}. Shifts by integer combinations of the vectors $\bs T^\parallel_{i}$ are exact symmetries of the potential \eqref{nppotentialtheta}, while  shifts by integer combinations of the vectors $\bs T^\perp_{j}$ break the periodicity, but by the least amount possible.

We refer to theories in which the relative angles between the constraint surface and the aligned basis are small, $\lVert \mathbfcal P\bs T^\perp_{j} \lVert_\infty \ll 1$, as {\it well-aligned}. In addition to the $N$ exact symmetries, well-aligned theories have $P-N$ \emph{approximate shift symmetries} generated by the  vectors $\bs T^\perp_{j}$.    

When $N \gg 1$  the determinant of $\mathbf Q$ is  large, so the denominators of the rational numbers appearing in first $N$ columns of $\mathbfcal R$ are  typically large.  If these numbers were irrational the angles would be arbitrarily small, and hence generically all the angles are   small when $N \gg 1$ and $P$ is not too large.  In general, Minkowski's theorem provides an upper bound on the smallest angle \cite{minkowski1911}. We have verified numerically that  theories with large $N$ are generically very well-aligned so long as $P$ is somewhat smaller than $2 N$.

The phases $\bs \delta$ in (\ref{phipot}) indicate the relative offset of the origin of the lattice basis from the constraint surface, and deserve some discussion. Clearly we can eliminate $N$  of the $P$ phases by continuous shifts of the axions. Furthermore we can perform discrete shifts of  $\bs \phi$ to set the $P-N$ remaining phases to zero within some finite accuracy. This corresponds to choosing the lattice point closest to the constraint surface to be the origin of the lattice basis. Hence, in well-aligned theories we can reduce the phases to zero within good accuracy.  In fact, small phases are necessary for the lattice and kinetic alignment mechanisms of \cite{Kim:2004rp, Bachlechner:2014hsa},  therefore our new effect justifies these previously discussed varieties of axion alignment.

Let us apply this technology to find vacua. Wherever the constraint surface is close to the center of a fundamental domain, a quadratic expansion of the auxiliary potential yields the vacuum locations
\be\label{location}
\bs\phi_{\text{vac},\,\bs m} = 2\pi (\mathbbold{1}_{P}-\bs \Delta)\bs T^\perp\bs m+\mathcal{O}(\bs\Delta \bs T^\perp\bs m)^2\,.
\ee
Here 
$$\bs \Delta=\diag{(\Lambda_I^{-1})}\mathbfcal R^\top(\mathbfcal R \,\diag{(\Lambda_I^{-1})}\mathbfcal R^\top)^{-1}\mathbfcal R,$$
  $\bs m$ is an integer $(P-N)$-vector, and $ \mathbbold{1}_{P}-\bs \Delta$ is the (generally non-orthogonal) projector onto the constraint surface which yields approximate vacua of the on-shell potential. The quadratic approximation is valid as long as the vacuum is well inside a fundamental domain.

Vacua outside the region of validity of the quadratic expansion can be found by numerically minimizing (\ref{phipot}). In general this is very time-consuming for exponentially large numbers of vacua, but the tools developed above  allow us to  overcome this difficulty. We can select a small but representative set of all the vacua by sampling fundamental domains that have non-vanishing overlap with the constraint surface, while ensuring that the domains in the sample are not related by shift symmetries (approximate or exact).
  Hence, our approach allows for a reliable statistical analysis of theories that may have vastly more vacua than could conceivably be sampled individually.

Let us illustrate the power of our approach with the simplest example of equal scales $\Lambda_I=\Lambda$ and $P=N+1$ nonperturbative terms.  Assuming that the entries of $\mathbfcal Q$ are independent and identically distributed we typically find  $| \bs \Delta \bs T^\perp | \sim |\det \mathbf Q \,|^{-1}$. This allows us to determine the vacuum energies in the quadratic approximation
\be\label{vacuumenergy}
V_{\text{vac},\,m}\approx{1\over 2}{\Lambda^4}\left( {2\pi\, m \over |\det \mathbf Q \,|}\right)^2+V_0\,,~~m\in \mathbb Z\,.
\ee
The quadratic expansion is valid well within the fundamental domains (\ref{funddomain}), which gives an estimate of the number of vacua $m\lesssim |\det \mathbf Q \,|$. At large $N$ the determinant of $\mathbf Q$ becomes extremely large, $|\det{ \mathbf Q}\,|\approx\sigma_{\Q}^{N}\sqrt{N!}$, where $\sigma_{\Q}$ denotes the r.m.s. value of the entries of $\mathbf Q$ \cite{goodman1963}. These estimates are valid in the universal regime where at least a fraction $\simgeq 3/N$ of the charge matrix entries are non-vanishing \cite{wood2012,Bachlechner:2014gfa}. Therefore, even at moderately large $N$ we find a vast number of minima whose locations and vacuum energy densities are given by (\ref{location}) and (\ref{vacuumenergy}). 
For example, with $N  = 200$ and $\sigma_\Q^2=2/3$ we  easily identified  $10^{165}$  distinct vacua on a desktop computer.  If  $V_{0} \sim - \Lambda^{4}$ this includes many minima with energies consistent with the observed vacuum energy of our universe \cite{Bousso:2000xa}.

The highest vacuum  in this simple example has an energy density $V_{\text{vac},\,\text{max}}\approx 0.14\times P\Lambda^4$,  well below the mean of the potential $P\Lambda^4$. Within the quadratic approximation, the vacua are distributed as $ 1/\sqrt{V}$, which yields a median vacuum energy density of roughly $V_{\text{vac},\,\text{max}}/4$. Neighboring vacua are easily identified in the lattice basis, typically lie at very different levels, and are separated by potential barriers of height $\gtrsim\Lambda$.

When the number of large nonperturbative effects  exceeds twice the number of axions, alignment typically fails. However, in this regime we can switch to an approximate description in terms of an isotropic Gaussian random field whose correlation functions match that of the axion potential. Again, typical vacuum energies are well below the mean potential \cite{Bachlechner:2014rqa,bk}.

\section{Vacuum Transitions}
Vacuum transitions are important for at least two reasons.  First, we should check whether they  destabilize  typical minima -- that is, whether the decay rate is faster than the Hubble rate.   This condition is most stringent for vacua with very low vacuum energy  (such as those compatible with our universe).  Second, eternal inflation in a meta-stable minimum with relatively large vacuum energy is a compelling choice for the initial condition and/or generic state of our universe (for instance  \cite{Linde:1983mx,PhysRevD.37.888,BDEMtoappear}). Tunneling from such a minimum naturally sets up initial conditions for inflation \cite{Freivogel:2005vv}.  If a sufficient number of efolds of slow-roll inflation follows and the field trajectory ends in a minimum with very small vacuum energy --- both of which are possible in these theories, and both of which are required by the criterion that galaxies are not exponentially rare  \cite{1981RSPSA.377..147D,Weinberg:1987dv,Freivogel:2005vv,Bachlechner:2016mtp} --- this would account for much of the expansion history of our universe.

The vacuum decay rate between vacua $\cal A$ and $\cal B$ in the thin-wall, no gravity approximation is given by $\Gamma_{\cal BA} \sim e^{-|B|}$, where
\be
B={27\pi^2 \sigma^4\over 2 (V_{\cal B}-V_{\cal A})^3}\,
\ee
and $\sigma$ is the tension of the wall.  A sufficient condition for  stability on gigayear time scales is very roughly $ \Gamma_{\text{total}} < 10^{-10^3}$, where $\Gamma_{\text{total}}$ includes a sum over $\sim 3^P$ neighboring vacua. To estimate the decay rate we need to take the kinetic terms in \eqref{lagtheta} into account.   The kinetic matrix in the lattice basis is given by $(\mathbf Q^{-1})^\top \mathbf K \mathbf Q^{-1}$ and has positive eigenvalues $\xi_N^2\ge \dots\ge\xi_1^2$.   The tunneling path in generic theories is no shorter than $\xi_N/\sqrt{N}$, and so this provides a weak bound on the parameters of the theory: 
$$\Lambda^4\ll 10^3 {\xi_N^4 \over P N^2} \, .$$ In the interesting parameter regime (e.g.~$P N^{2} \sim 10^{6}$, $\xi_{N} \sim \M,\, \Lambda \sim 10^{-2} \M$)  a vast number of vacua are extremely stable (see also \cite{Masoumi:2016eqo}).

\begin{figure}[t]
\centering
\includegraphics[width=.48\textwidth]{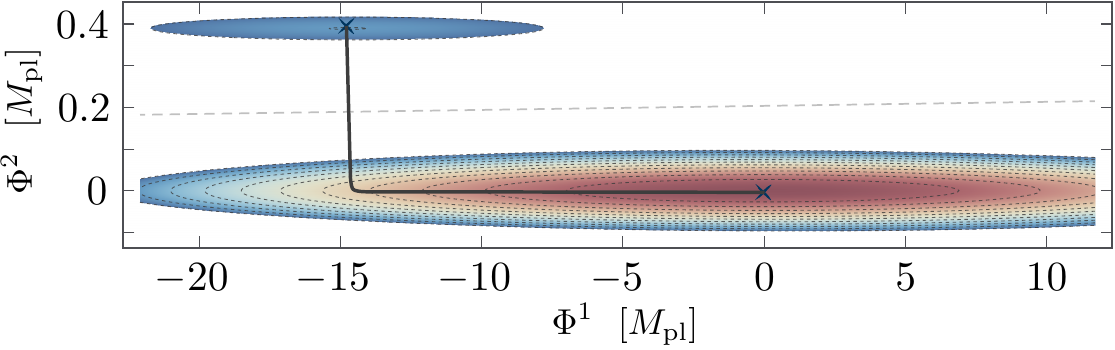}
\caption{\small Equal potential contours over a two-dimensional slice containing two vacua. The dashed line denotes the boundary of the fundamental domain. This  allows for Coleman-de Luccia tunneling from a false vacuum to an inflationary slope.}\label{fig:tunnel}
\end{figure}

In general there  is a  tension between thin-wall vacuum decays that require $V''\gg H^2$, and a sustained period of inflation that demands $V''\ll H^2$ \cite{PhysRevD.59.023503}.  In the multi-axion models at hand, however, the  hierarchies in $\xi_i$ render the canonical field space highly anisotropic, and these theories generically accommodate thin-wall vacuum transitions in bubbles that subsequently inflate in an open FLRW cosmology. We give an example in Figure \ref{fig:tunnel} (note that the kinetic energy gained by the inflaton after tunneling does \emph{not} cause the field to overshoot the inflationary plateau, for reasons explained in \cite{Freivogel:2005vv}).

\section{Inflation}

The implementation of inflation purely within effective field theories is famously tentative at best: physics above the cutoff can spoil an inflationary trajectory or destabilize the theory altogether. Despite this limitation it still is instructive to consider the cosmological observables that would arise in the absence of any such effects.

Shift symmetries provide for some of the most compelling, radiatively stable models of inflation  \cite{Natural}, that can be embedded in string theory \cite{Baumann:2014nda}. Most notable are variants of assisted inflation, that exploit multiple axion shift symmetries to ensure the super-Planckian field space diameters required for large-field inflation, ${\cal D}_{\text{inf}}\gtrsim\M$ \cite{Liddle:1998jc,Nflation,KNP}. While the invariant diameters for single axions are  sub-Planckian when the perturbative expansion is well-controlled \cite{Banks:2003es}, axions are numerous in typical flux compactifications and generically allow for collective field space diameters that significantly exceed the ranges of the individual axions via kinetic and lattice alignment \cite{Bachlechner:2014gfa,Long:2016jvd}.  In lattice coordinates, the relevant boundaries of typical fundamental domains roughly correspond to an $N$-hypercube that has no special orientation with respect to the least massive axion, which thus is well-aligned with one of the numerous diagonals. This phenomenon of kinetic alignment generically yields an axion diameter as large as \cite{Bachlechner:2014gfa,Bachlechner:2014hsa} 
\be \label{infrad}
{\cal D}_{\text{inf}}\sim 2\pi \sqrt{N} \xi_N\,,
\ee
which can exceed the field ranges of each individual axion by far. 

We can comprehensively sample the inflationary dynamics by marginalizing over the charges ${\mathbfcal Q}$ and phases $\bs \delta$ of the instantons, the kinetic matrix $\bs K$ and the constant energy density $V_0$. Since we are only interested in vacua that satisfy the selection bias constraint of (almost) vanishing cosmological constant we can marginalize over $V_0$ by considering only those values that ensure a vanishing energy density at the vacuum reached at the end of inflation. The tools developed above provide us with a representative sample of all vacua, so by considering all initial conditions that can terminate in that vacuum we find a representative collection of all possible cosmological histories.

We study the classical dynamics by solving the equations of motion for the fields and the Friedmann equation for the scale factor of a homogeneous FLRW cosmology,  discarding any solution inconsistent with our selection bias \cite{GrootNibbelink:2001qt,Price:2014xpa}. Whenever the single field, slow-roll approximation is valid throughout the evolution \cite{Wands:2000dp}, the scale of inflation, tensor-to-scalar ratio and spectral index respectively are given by
\be
E_{\text{inf}}\approx 0.01\times r^{1/4}\M\,,~~~r\approx 16\epsilon\,,~~~n_{\text{s}}\approx 1-2\eta-4\epsilon\,,
\ee
where $\epsilon=-\dot{H}/H^2$ is the first, and $\eta=\ddot{\bs \Phi} \cdot \bs e_\parallel/|\dot{ \bs \Phi} |H$ is the second slow-roll parameter projected onto the  tangent of the trajectory $\bs e_\parallel$.  All quantities should be evaluated at horizon exit.  
The single field approximation is valid when the acceleration transverse to the field velocity and non-adiabatic particle and/or string interactions are negligible \cite{Chen:2009zp, Green:2009ds, DAmico:2012wal, Assassi:2013gxa,Flauger:2016idt}.

To study inflation in our model we sample the trajectories leading into each vacuum, with $V_{0}$ chosen in each case so that that  vacuum has zero energy.  We discard any trajectories with less than 60 efolds of inflation.  We find two qualitatively different regimes. Whenever inflation proceeds over a super-Planckian distance within one single fundamental domain, as is the case in the aligned axion inflation scenario discussed in \cite{Bachlechner:2014hsa,Bachlechner:2014gfa,Kim:2004rp,Dimopoulos:2005ac}, we find a lower bound of $r\gtrsim 0.07$, and the single field approximation is valid. In generic theories (assuming roughly constant $\Lambda_I$), the vacua are very low compared to the mean of the axion potential and therefore downward vacuum transitions can only source this regime. 

However, when inflation proceeds at typical scales of the potential, much more diverse  features in the potential are encountered, such as hill-tops and saddle points (see also \cite{HT1,ht2,Czerny:2014xja,Czerny:2014qqa}). Even in very simple theories we observed a wide range of $n_{\text{s}}$ and values of $r$ as low as $r\sim 10^{-4}$, but we speculate that much lower values of $r$ are possible in more complex models. The single field approximation breaks down for some trajectories and it is not  clear whether non-adiabatic perturbations decay by the end of inflation to allow for a simple treatment. These results motivate a future study of the multi-field dynamics and perturbations. The corresponding initial conditions can  be sourced by upward  transitions or other mechanisms (e.g.~\cite{Dienes:2015bka, Dienes:2016zfr}). 
Note that while $n_{\text{s}}$ and $r$ are independent of the overall scale of the potential, the power spectrum depends on the scale, and so to match observation, trajectories with smaller $r$ must occur in models with correspondingly smaller values for the $\Lambda_{I}$.

A single multi-axion theory can accommodate a very diverse set of inflationary trajectories with significantly different cosmological observables. This finding highlights the necessity of a detailed understanding of inflationary initial conditions to satisfy  even the most basic prerequisites for definite predictions in multi-axion theories.

\section{Dark Matter}
\label{dmsec}

If  ${\mathbfcal Q}$ is not full rank, the leading potential  \eqref{nppotentialtheta} leaves at least one of the fields -- $\Phi_{\text{light}}$ -- with an unbroken continuous shift symmetry. It is generally believed that theories of quantum gravity do not permit  continuous global symmetries, so there should exist a  subleading instanton with action $S_{G}$ that breaks the continuous shift symmetry by a term $V_{G} \sim \M^{4} e^{-S_{G}} \cos(\Phi_{\text{light}} /f_{\text{light}})$. A typical axion decay constant of the leading nonperturbative term is   $f_{\text{light}} \approx\sqrt{\pi/8} \, f_N$, where $f_N^2$ denotes the largest eigenvalue of the field space metric $\bs K$ \cite{Bachlechner:2015qja}.  A natural guess for $S_{G}$ is provided by the weak gravity conjecture\footnote{The weak gravity conjecture in general is not inconsistent with large field inflation, see e.g. \cite{Cheung:2014vva,Rudelius:2015xta,Madrid,Madison,Brown:2015lia,Heidenreich:2015wga,Heidenreich:2015nta,Saraswat:2016eaz,Hebecker:2017wsu,McAllister:2016vzi}.}, which asserts that no gauge interaction is weaker than gravity \cite{ArkaniHamed:2006dz}. Extending this conjecture to axions provides an upper bound on the action, $S_{G} \lesssim \M/f_N$ \cite{Cheung:2014vva,Rudelius:2015xta}. The bound is approximately saturated by euclidean wormholes  that couple to $N$ axions, which here yields $S_G\approx\sqrt{3\pi N}\M/2f_N$ \cite{ArkaniHamed:2007js,Madrid,Bachlechner:2015qja}.  This allows us to estimate the mass of the lightest axion:
 $$
 m_{\text{light}} \approx {\M^{2} \over  f_{\text{light}}} e^{-S_G/2}.
 $$
A mass of roughly $10^{-22} \, \text{eV}$   has the  virtue that it ameliorates the problems conventional CDM models have at sub-kpc scales by suppressing structure below the Compton wavelength $m_{\text{light}}^{-1}$ \cite{Hu:2000ke, Arvanitaki:2009fg, Hui:2016ltb}.   Choosing $N \sim 100$ and  $f_{\text{light}} \approx  .04 \, \M$ yields $m_\text{light} \approx 10^{-22} \, \text{eV}$.   Remarkably, with these numbers axion misalignment  
 generates roughly the correct dark matter abundance:
\be
\Omega_{\text{axion}} \sim 0.2 \left({ f_{\text{light}} \over .04  \,\M}\right)^{2}  \left({m_{\text{light}} \over 10^{-22} \, \text{eV}} \right)^{1/2}.
\ee

The eigenvalue $f_N \sim f_{\text{light}}$ is typically related to $\xi_{N}$ by $\xi_{N} \simleq N f_{N}$.  The typical diameter of the fundamental domain is roughly ${\cal D_{\text{inf}}}\lesssim 2\pi N^{1/2} \xi_N$  \eqref{infrad} \cite{Bachlechner:2014gfa}. Hence, with $N\approx 100$ multi-axion models can accommodate both Planckian field space diameters ${\cal D_{\text{inf}}}\sim \M$ and light axions that reproduce the observed dark matter abundance $\Omega_{\text{axion}}\sim 0.1$.

It is something of a miracle that this analysis gives parameters with the correct range of values to both give the correct dark matter abundance and help solve this problem (see also \cite{Arvanitaki:2009fg, Hui:2016ltb}). 
 Given this surprising observation one may be led to speculate that dark matter could be related to nonperturbative gravitational physics.
 
 The combination of this ``fuzzy'' dark matter and high-scale inflation can lead to an overproduction of isocurvature modes.  Avoiding this probably requires  $r \simleq \text{few} \times 10^{-4}$  (see for instance \cite{Marsh:2015xka,Diez-Tejedor:2017ivd}). Values of $r$ this small appear for inflationary trajectories even in the simple $P = N+1$ model discussed above, and we expect larger $P-N$ to provide even more diversity.

\section{Acknowledgements}
We would like to thank Lam Hui, Nemanja Kaloper, David J.E. Marsh, Liam McAllister and Alberto Tejedor for useful discussions. We thank Eva Silverstein for suggesting the title. The work of TB and KE was supported by DOE under grant no. DE-SC0011941.  The work of MK is supported in part by the NSF through grant PHY-1214302, and he acknowledges membership at the NYU-ECNU Joint Physics Research Institute in Shanghai.

\bibliographystyle{klebphys2}
\bibliography{tunnelingrefs.bib}

\end{document}